\documentclass[letterpaper, 10 pt, conference]{ieeeconf}  

\IEEEoverridecommandlockouts                              

\usepackage{soul}
\usepackage{booktabs}
\title{\LARGE \bf
A blessing or a burden? Exploring worker perspectives of using a social robot in a church
}

\author{Andrew Blair$^{1}$, Peggy Gregory$^{1}$ and Mary Ellen Foster$^{1}$
\thanks{$^{1}$Andrew Blair, Peggy Gregory and Mary Ellen Foster are with the School of Computing Science,
        University of Glasgow, United Kingdom. Correspondence address:
        {\tt\small andrew.blair@glasgow.ac.uk}}%
}

\begin{document}

\maketitle
\thispagestyle{empty}
\pagestyle{empty}

\begin{abstract}
Recent technological advances have allowed robots to assist in the service sector, and consequently accelerate job and sector transformation. Less attention has been paid to the use of robots in real-world organisations where social benefits, as opposed to profits, are the primary motivator. To explore these opportunities, we have partnered with a working church and visitor attraction. We conducted interviews with 15 participants from a range of stakeholder groups within the church to understand worker perspectives of introducing a social robot to the church and analysed the results using reflexive thematic analysis. Findings indicate mixed responses to the use of a robot, with participants highlighting the empathetic responsibility the church has towards people and the potential for unintended consequences. However, information provision and alleviation of menial or mundane tasks were identified as potential use cases. This highlights the need to consider not only the financial aspects of robot introduction, but also how social and intangible values shape what roles a robot should take on within an organisation.
\end{abstract}

\section{Introduction}
Robots have altered the workplace since the 1960s, when the first programmable industrial robot arm was deployed into an assembly line \cite{gasparetto_unimate_2019}. Since then, many industrial and service robots have been developed for the manufacturing sector, with line-following robots moving materials throughout factories and robotic arms performing complex tasks such as spray painting or spot welding \cite{engelberger_robotics_2012}. 

Until recently, real-world robots were deployed almost exclusively in the industrial and manufacturing sectors due to their controlled environments. However, advances in localisation,  mapping and algorithms have enabled autonomous robots to operate reliably in less predictable working environments \cite{abaspur_kazerouni_survey_2022}. This has resulted in the service sector being able to integrate robots into their workflows: for example, hospitality staff can now be assisted to serve customers in restaurants \cite{el-said_are_2022}, and home deliveries can be completed autonomously \cite{pelikan_encountering_2024}.

Traditionally, robots assist with manual, physical tasks. However, with the introduction of end-to-end transformer-based automatic speech recognition \cite{radford_robust_2022}, large language models \cite{minaee_large_2024} and advancements in social signal processing, tasks and interactions reliant on conversation can now also be supported by robots. This means that effects on job transformation and displacement can now additionally be seen within information provision roles \cite{rukumnuaykit_post-covid-19_2024}.

This extension to knowledge-based domains has resulted in social robots being seen in public spaces such as airports \cite{hwang_difference_2024}, shopping centres \cite{agah_mummer_2016}, and train stations \cite{lajante_when_2023}. These public spaces can be categorised as primarily \textit{functional} spaces, in which the focus is on transactional interactions that help the user achieve goals, such as guiding them to their flight or purchasing goods. Within these spaces, robot introduction could be seen through a purely capitalist lens, with deployment decisions based on a cost-benefit analysis comparing robot to human labour \cite{decker_service_2017}.  However, if profit and efficiency are not the only motivators for an organisation, such as 
in the public or third sector, there are additional competing factors influencing robot introduction. 

These sort of deployment contexts are generally under-explored, although two sub-sectors have seen some work. Social robots have been seen within university buildings \cite{blair_development_2023}, often benefitting from the innovative culture and openness to novel experiences. Museums, who often have to balance profits with creating an experience for a visitor, have also explored social robots as a means of engagement \cite{duchetto_lindsey_2019,thrun_minerva_1999,rodriguez_personal_2020}. However, these robots are most often deployed within science museums \cite{gasteiger_deploying_2021}, with the robot being an attraction in itself rather than necessarily delivering value as an assistant in its own right.

A more overlooked context, with an even more diverse set of motivators, is that of churches: they provide spiritual and emotional support, assist with educational opportunities, undertake advocacy work and host community events \cite{sider_churches_2002}. This is underpinned by a strong focus on community, with Dietrich Bonhoeffer famously stating \textit{``The Church is the Church only when it exists for others''} \cite{bonhoeffer_letters_2010}. These non-monetary values and social benefits are considerably more challenging to assess, and they present a more complex challenge for robot introduction than for functional spaces like airports. The nature of a church creates a unique blend of competing stakeholder priorities and values, and  to ensure a successful robot deployment, these values must be explored and considered with church workers to produce a system fit for everyone's needs. 


To address this under-researched area, we have partnered with a large working church and visitor attraction in the United Kingdom to explore the potential roles of a social robot within the church. The church attracts a wide range of people, with some viewing the space as a centuries-old spiritual sanctuary and others seeing it as a place of architectural beauty with great photo opportunities. Total visitor numbers are increasing year on year, putting growing pressure on an understaffed organisation. They are currently looking for ways to decrease this pressure, and one potential solution is the use of social robots. In this paper, we explore how workers within the church perceive and respond to the proposed introduction of a social robot. We seek to answer the following research questions:


\begin{itemize}
    \item RQ1: How do staff respond to the idea of a social robot being introduced to the church?
    \item RQ2: What are the competing human perspectives from staff about the tasks the robot should do?
    \item RQ3: How is robot introduction affected by competing values 
    in value-rich spaces?

\end{itemize}
\section{Background}
\label{related}

This paper seeks to contribute to and extend existing research by combining two under-explored areas. We elicit perspectives about introducing a social robot from a diverse set of workers, and do so within the unique deployment context of a spiritual institution.

\subsection{Robots within spiritual institutions}
Automation within spiritual institutions has a long history, with the first vending machine in the form of a coin-operated automata to dispense holy water being placed within a temple in 215 BC, to alleviate the demands on priests \cite{obrien_exploring_2021}. Automata could commonly be found in Catholic churches in the medieval period, often as additions to mechanised clocks and representing clergy or religious figures \cite{riskin2010machines}. However, robots that depict social agents within a spiritual or religious context have largely been confined to theoretical work \cite{trovato_introducing_2016}. Limited social robot deployments have been trialled, with various platforms being used to provide remote attendance to funerals \cite{arnold_cybernetic_2021}, as a tool to disseminate homilies and bible verses \cite{trovato_communicating_2019}, and to give blessings to visitors \cite{loffler_blessing_2021}. 

\subsection{Diverse Worker Engagement}
Traditionally when exploring real-world deployment contexts researchers focus solely on end-user perspectives towards robot introduction \cite{martinez_hey_2023,nakanishi_robot-mediated_2022,donnermann_integrating_2020}, but recently some researchers have investigated worker perspectives and impacts on organisational dynamics.
The Eurofound research project, exploring human-robot workplace changes \cite{riso_human-robot_2024}, found productivity gains were the main reason for adopting robots. Additionally, Eurofound noted that employees were not actively involved in the decision-making process, with the responsibility for adopting the robots resting solely with top-level management. The benefits of participatory approaches when adopting workplace technology are well understood \cite{ehn_work-oriented_1988}, yet we can see that in practice they are often ignored.
The first workshop on \textit{``Worker-Robot Relations''} in 2024 explored these challenges \cite{zaga_first_2024}. Topics covered including front-line workers as primary stakeholders \cite{foster_including_2024}, addressing knowledge hiding \cite{alabak_2024} and attachments that can form between robots and workers \cite{mollen_2024}.

\section{Methodology}
This paper presents one part of a larger design study delivered in collaboration with the church. We focus here on the interviews we undertook with staff members. 
We were granted ethical approval to carry out this study from the University of Glasgow College of Science and Engineering Ethics Committee (\#300240015).
\subsection{Stakeholder Identification}
We first identified the five stakeholder groups within the church who were part of the workforce. Predominantly the front-line workforce is made up of \textit{volunteers}, who give unpaid time to answering questions, leading tours and talking to visitors. There is are also paid members of \textit{front-line staff}  present daily. There is also a \textit{heritage management team} who are concerned with disseminating the history and cultural aspects of the church and are responsible for providing the front-line experience for visitors. There is a \textit{clerical team} that manages the spiritual aspects of the church, such as services and life events, as well as national and ceremonial events. This includes the ministers who lead the church and congregation. Finally, there are various \textit{back-office staff} who manage tasks such as finances and administration.

\subsection{Participants and Procedure}
Fifteen participants were recruited from the five stakeholder groups identified above. This was achieved through a recruitment email, a poster placed on the staff noticeboard, and word of mouth. Participation in this study was completely voluntary and did not affect anyone's continued role in the church. No payments were made for their participation in the study. To preserve the anonymity of participants, we only report the stakeholder group that they belong to; this is due to the small size of the organisation. 

Five participants were volunteers, two were paid front-line workers, three were from heritage management, three were clerical staff and two were other staff members. Participant ages ranged from the 25-34 group to 65+. Almost all participants had no prior experience with robots, apart from one member of heritage management having seen a demonstration of one at a university open day and a volunteer taking part in a university lab study. 

We conducted a semi-structured interview with each participant lasting approximately one hour. The interviews took place within various meeting rooms in the church itself, with one being hosted online on Microsoft Teams. Before the interview, participants were asked to fill in a consent form and provide basic demographic information: age group, job title and experience with robots.

Interview questions covered three topics: 1) the interviewee's job role, 2) the purpose of the church and 3) the robot. 
Interview questions were tailored depending on the stakeholder group: for example, managers were asked how they trained staff, while employees were asked about their training experience. We employed laddering theory \cite{reynolds_laddering_2001}, where participant responses are continually probed with ``why'', to try and understand underlying motivations.

\subsection{Data analysis}
The interviews were transcribed and used as the data for our analysis. We used a six step approach to reflexive thematic analysis as defined by Braune and Clarke \cite{braun_using_2006} to explore and understand the data, and the first author coded the transcripts inductively. This approach was chosen as we are researching a novel and under-explored area, and we wanted our themes to be grounded in the data.

\subsection{Reflexivity and Validity}
\label{quali-inquiry}
The first author carried out the interviews and the subsequent coding of the transcripts. They have no affiliation with the church the study was completed in, are not a trained heritage professional, and do not have a background within the clerical or wider body of the church. Participation in our study was unpaid and voluntary, and we were not able to interview every volunteer and member of staff at the church. It is likely that we were therefore not presented with more negative viewpoints towards the robot's introduction due to participation bias. This is difficult to mitigate especially with volunteers, as they are giving up their own personal time to take part in this study.

We address the four criteria for trustworthiness in qualitative inquiry defined by Lincoln and Guba \cite{lincoln_naturalistic_1985}. We establish \textit{credibility} through triangulation of sources by interviewing participants from a range of stakeholder groups across the organisation, ensuring that participants are aware of their anonymity and conducting this study independently. \textit{Transferability} of our findings is limited due to only studying one organisation, but we provide rich descriptions of the research context while maintaining anonymity of the organisation. \textit{Dependability} is addressed through reporting our research methodology. \textit{Confirmability} is addressed by providing a reflexive statement and again our triangulation of sources.

\section{Results}


We identified four overarching themes from our reflexive inductive thematic analysis which we describe in detail below. In the quotes below, identify each participant by their stakeholder group using an acronym: paid front-line workers as FLS, volunteers as VLS, heritage management as HM, clerical staff as CLS, and other staff as OS. 

\subsection{The human element}

Participants expressed strong feelings about the role of the church within the community, and how the church facilitate this. VLS stated \textit{``of any institution, the church needs to be the most empathetic institution of all.''} This was supported by other participants noting the personal and intimate aspects of faith, with CLS commenting \textit{``I don't think people would like to speak about their dying grandmother to a robot.''} CLS suggested congregants would be \textit{``horrified if there was a robot giving a sermon,''} with CLS echoing these thoughts when considering an experience they had earlier in the week \textit{``[this] man that asked for prayers [from us]
... that's not a job for a machine.''}

This theme extends into the workforce, with participants stating their motivation for volunteering or working at the church comes from their belief that they are able to contribute to the experience. VLS stated \textit{``I feel I can offer something and I feel I can do it,''} while FLS stated the cultural environment of the church allowed a \textit{``great combination of I feel competent and I also feel valued.''}
Other participants stated their motivation was from a place of passion and interest for the church and its history, with VLS stating their volunteer role \textit{``fulfills my interest in history and in church history.''} OS suggested it was almost a pre-requisite for being part of the workforce, saying that \textit{``first and foremost be passionate about where you're working, about [the church].''}

Participants believed a key strength of the church was their ability to provide for a wide range of people, and that it was only possible to do this due to the diversity of the staff and volunteers. FLS stated \textit{``it's essential to have an element of variety ... because who comes here are visitors, who are infinitely varied [themselves],''} with OS saying about the volunteers \textit{``they have their own sorts of areas of interest, which I think comes across quite well in their tours and things.''} HM summarises these thoughts: \textit{``I think the variety is so important to how we work here.''} 

However, some participants did think the robot could be beneficial in offering an alternative engagement opportunity to visitors. FLS shared their experiences where \textit{``you often get people say, oh, this is a stupid question,''} and HM explored the idea that some people \textit{''might find [speaking to  a person] really intimidating and scary, but a robot is non-judgmental.''}
HM considered this too, saying \textit{``people who want to speak to something, but are really nervous around humans. Could a robot fulfill that gap for them?''}
This displays how some participants feel that the robot could help the church connect more with people they currently struggle to reach, but with clear limitations and the feeling that only a human is suitable for many aspects of the church.

\subsection{Modernisation whilst preserving tradition}

Participants explored how the modernisation of the building could improve and enhance the experience for everyone. OS explained that a new lighting system that had been installed due to the existing one not being \textit{``fit for purpose,''} while VLS highlighted that the church was the first in the country to introduce contactless terminals as a means of donation. HM spoke about prior work to develop an interactive video game to disseminate the history of the church, saying \textit{``I've heard very little negative, except that it looks a bit, you know, it looks like a screen in a building.''} Staff spoke of struggles with visitors respecting traditions with their own technology, with VLS describing a situation on Maundy Thursday where they had to intervene as a visitor was \textit{``using their telephone right beside the holy table.''} This response to visible and present technology was echoed by some staff, with OS stating \textit{``I don't enjoy having the masses of TVs everywhere, there's something that looks [wrong with them being] wheeled into a hundreds of years old building,''} and CLS saying \textit{``the technology we have shouldn't be obvious.''} CLS also highlighted that some congregants shared this aversion to television screens, and thought that the introduction of a social robot would be met with an even stronger response. Some participants explored this idea further, discussing the difference between how people accepted technology that silently enhanced existing functions versus a visible technology designed to do something new. VLS stated they believed people thought \textit{``we need that because ... you make it more accessible. Whereas a robot ... it's staring them in the face and challenging them.''}

Some participants worried that the robot could disrupt the space, and possibly take away from the experience and feelings that people may get from the church. HM suggested visitors may think \textit{``Actually, who cares about the chapel? There's a robot in here.''}. HM shared these concerns, and thought when people entered one of the side chapels \textit{``the sense of inherent awe when you go in there could be thought to be lessened by there being a robot in there.''} They also pointed out that \textit{``when [the architect] designed the [side chapel they] didn't design it thinking at one point it would house a robot,''} reflecting again how a visible addition to the building could alter the visitor experience.

The introduction of a robot was seen by some as a potential to showcase the church as a modern organisation. VLS wondered \textit{``can the robot be used as a means of basically, particularly younger people, sending a message that the church is old but its thinking is modern?''} However this was not an ubiquitous view, with CLS expressing concerns that \textit{``some of the elders and the members of the congregation still don't like signage up, so getting a robot involved [could be difficult].''} Some participants also expressed concerns around the robot being seen as an unserious or novelty item, with HM explaining \textit{``you can kind of picture it ... everyone just like, [the church] does robots.''} CLS considered the wider public perception, sharing their fear that the church would be \textit{``mocked in the press for either being gimmicky or putting a machine where there should be a person.''} Participants highlight through this theme the complex intangible values that must be navigated when considering the robot introduction; modernisation of the church is necessary but it cannot be at the expense of the core traditional values of the church.

\subsection{Improving experience}

This theme was expressed in two related ways: participants identified that a robot might be able to address the need for staff to carry out menial or mundane tasks, and also suggested that a robot might be a useful addition to the workforce alongside the human staff.

\subsubsection{Alleviating menial or mundane tasks}
There were clear tasks identified by participants for which they would welcome robotic assistance.  The space often requires reconfiguration for different events multiple times throughout the week, with FLS saying \textit{``design us a nice chair mover, that would be wonderful''} and HM recognising this too: \textit{``I think the staff would love like a chair moving robot.''}

Repetitive tasks were mentioned by participants as of particular interest for robot assistance. One specific location that was identified by front-line staff was the welcome desk. They spoke of having to repeatedly stop visitors with drinks entering the building, and that they would get tired from the repetitive nature of the interaction, with FLS describing it as \textit{``almost like a fatigue.''}  Some participants found the welcome desk a less engaging or rewarding part of their role compared to tours or other visitor interaction, with FLS explaining they were \textit{``going to have to be stuck on the desk''} after the interview.
However, some participants saw the robot at the welcome desk as a poor idea. CLS voiced their concerns from the point of view of visitors, saying \textit{``It shouldn't be in the welcome desk, saying welcome to [the church]. I think you want somebody from [the church], a person, saying welcome.''} VLS voiced concerns about the effect it could have on some volunteers' motivation, explaining \textit{``[the welcome desk] is a bit about my job I don't really enjoy ... but [the robot replacing the welcome desk role] wouldn't do somebody who really likes doing that [role].''}

There are various security aspects to the roles that participants mentioned as not being desirable. Front-line staff spoke of having to intervene with visitors entering restricted spaces, and this constant monitoring of the space VLS described as \textit{``detrimental to the visitor experience. You're trying to focus on answering their questions and at the same time you're having to try and have one eye on what's going on elsewhere.''} 

Understaffing was also highlighted as a problem within the church, with HM simply stating \textit{``We need more staff.''} VLS suggested one of the knock-on effects of this was that experienced front-line staff were therefore not utilised effectively, saying \textit{``at the moment, [my colleague] is on the desk and what a waste of [their] skill set.''} 

Overall, the FLS summarised this theme well: \textit{``[if the robot] could sort of take some of the drudgery out of the job while still allowing me to do the stuff I do enjoy, then yeah, I certainly wouldn't be complaining.''}

\subsubsection{Enhancement and extension of human potential}
The robot was also identified as being able to provide additional help to front-line staff to increase engagement within the church. Some participants considered it an opportunity to standardise information sharing, with VLS suggesting 
it could \textit{``help to provide consistency''} to the visitor experience. 

Whilst participants spoke of alleviating tasks considered mundane or menial, they also saw the proposed introduction of the robot as a way to extend and augment their own personal capabilities. HM stated they saw the robot's purpose \textit{``to fill gaps, but also fill gaps that can't be filled with any other solution.''}
One such augmentation the robot could provide that was almost universally identified by participants was translation. Volunteers highlighted their experience of language barriers when interacting with visitors, with VLS stating \textit{``I wish I was a multilingual person.''} HM thought \textit{``the ability to access information in a whole host of languages would be amazing and that would make a big impact to the visitor's experience.''} We can see that participants could see the robot as beneficial to both the visitors and to their own work, or that of their staff.

\subsection{Robot agency}
Many staff expressed concerns about the robot's agency: that is, how the robot would make decisions about what to say and do. One frequent area of discussion surrounded the knowledge base of the robot, with VLS wondering \textit{``where the cut-off point is in terms of dialogue and discussion, and who controls that''} and VLS indicating they didn't \textit{``want to bring the size of a planet\footnote{In reference to the brain of science fiction character Marvin the Paranoid Android, from the radio series ``Hitchhiker's Guide to the Galaxy'' \cite{hitchhikers}}
onto the front desk.''} HM believed the robot \textit{``should stay on topic, and the topic we want it to stay on is heritage.''} VLS agreed with this, saying they would want it \textit{``limited to answering questions about [the church]''}. However, note this was in contrast to what staff currently do, with VLS also stating \textit{``the people on that desk do actually answer questions on multifarious subjects that have nothing to do with [the church], for example where is [external attraction].''}

Subjectivity was also expressed as a concern. HM felt that the robot \textit{``shouldn't try and talk about religion, or certainly the subjective bits of religion.''} VLS highlighted the potential for offence to be caused from prolonged interactions, saying \textit{``you don't know how someone's going to take it, or it leads to bigger conversations ... it's not fair for the little robot.''} This collectively suggests that participants believe a robot may need to be held to different or higher standards compared to a human member of staff.


\section{Discussion}
We have explored worker perspectives to the proposed introduction of a social robot into their church. We found that worker perspectives could be broadly categorised into four themes: \textit{1) the human element}, \textit{2) modernisation while preserving tradition}, \textit{3) improving experience} and \textit{4) robot agency}. This section will now explore these themes through the lens of our three research questions:
    RQ1) How do staff respond to the idea of a social robot being introduced to the church?,
    RQ2) What are the competing human perspectives from staff about the tasks the robot should do? and
    RQ3) How is robot introduction affected by competing values in value-rich spaces?


As we can see through our thematic analysis, worker responses to the proposed social robot introduction were mixed. Many felt excited about the project, and saw it as a way to tackle some of the issues they have been facing surrounding understaffing within the church. It is however evident that the building itself invokes a lot of personal feelings, and the importance of this cannot be understated. 

This church embraces modernity, which is evidenced by its work to upgrade infrastructure systems and incorporate passive technologies such as induction loops. However, visible technology like television screens has previously met with resistance, and social robots may provoke an even greater response. Participants felt as if the robot could impair a visitor's perception of the space, and damage their take-home experience irreparably, even if it could provide useful knowledge to a visitor. It could also interfere with the spiritual aspects of the church, as people associate the church's core values with empathy and community, which are in direct conflict with the general perception of robots. This is not nearly as large a problem in a functional space; if the service quality is increased, then customer satisfaction is increased whether by human or robot service \cite{hwang_difference_2024}.

Participants universally recognised that physical tasks were time-consuming and hard work where the front-line staff would appreciate having robot assistance. Front-line staff identified the welcome desk as an often repetitive and tiring aspect of their role; they felt the robot could relieve them and instead allow them to staff other more rewarding areas of the church. However, management and the clerical team objected to the placement of the robot at the entrance, feeling it could provide a subpar, unwelcoming and even possibly offensive first introduction to visitors. This again links back to the wider perception of the church by the public, and highlights management concerns surrounding any reputational damage the robot could cause.

Some areas were less disputed by participants but at odds with the literature. Specifically, interviewees did not think the robot would be suitable for running the tours that the guides carry out. This was surprising at first, as robot tour guides are an extremely common deployment context 
\cite{duchetto_lindsey_2019,thrun_minerva_1999,rodriguez_personal_2020}. This feeling was expressed across stakeholder groups: the clerical team strongly stated they did not want the robot to engage users with any of the pastoral services they provide, with other participants pointing to how the robot should avoid religious discussions. This aversion to religious and spiritual interactions is at odds with the previous deployments we mentioned in Section \ref{related}, which even go as far as to lead users through intimate acts of faith such as praying \cite{trovato_communicating_2019}. 

The recurring issue that emerged was whether the robot would improve the \textit{visitor} experience or the \textit{worker} experience. Whilst understaffing and increasing visitor numbers can be seen as a motivator to ``automate the boring stuff,'' it is also an opportunity to try new engagement avenues with visitors. However, 
as the church is heavily reliant on volunteers, the management need to ensure the work provides benefit and value to the volunteer in order to retain them \cite{bekkers2016people}. If the church removes the work the volunteers enjoy, such as engagement near historic artefacts in the church, and instead makes them focus on more repetitive work like greeting, they risk losing volunteers and worsening their staffing shortage. We note that the management was aware of this risk, and allowed us access to interview their staff to help understand their concerns towards job transformations in the church.

One could therefore consider that robot deployments should be seen as a non-zero-sum scenario. This concept, originating within game theory and adopted by economic theory \cite{jaskiewicz_nonzero-sum_2018}, specifies that the reward to all participants is unfixed. This means that if the robot fails, or workers become alienated, or both simultaneously happen there would be a net negative outcome.

In the future, we aim to develop and deploy a robot within the church while considering the perspectives and values provided within these interviews. With suitable consideration of worker perspectives, and therefore the monetary, social and intangible values of the organisation, the risks of having both robot failure and wider organisational and reputational damage can be significantly mitigated. This approach increases the likelihood that robot integration is successful for workers and ultimately the organisation and its users.

\addtolength{\textheight}{-12cm}  
\end{document}